# The One-Zone Model of Cepheid's Pulsations by Zhevakin, Reconsidered

Mikhail I. Ivanov[*]

**ABSTRACT**

The one-zone model of non-adiabatic radial stellar pulsations is considered. Contrary to the original Zhevakin's work the triple-alpha process is the basic thermonuclear fusion reaction within the star under investigation. The model has four dimensionless numbers. For real Cepheids, the magnitude of each of these dimensionless numbers is estimated. It is found that dimensionless numbers associated with both radiation confinement and thermonuclear energy generation are sufficiently less than 1, that is, pure hydrodynamic effects prevail in the model studied. For this case, the model system reduces to the well-known equation of Bhatnagar and Kothari, which describes adiabatic radial pulsations. But solutions of this equation do not have many of real Cepheid curve's features (for instance, asymmetry of luminosity curves about mid-period points). Thus, the model under consideration is not satisfactory for describing Cepheid's pulsations. The problem of construction of an adequate simplified model of stellar pulsations is briefly discussed.

## INTRODUCTION

Stellar pulsation phenomenon is one of the most interesting problems of astrophysics. Observations show a wide variety of stellar pulsation types [1-3]. It should be thought that pulsation mechanisms are also different. In addition, one observes stars as though they are located between two seemingly isolated pulsation types (for example, V19 in M33 [4]) or stars showing temporal dynamics to be difficult to explain (for example, RU Cam [2, 5]).

Nowadays stellar pulsations are usually studied by numerical modelling by standard pulsation codes. But some important questions about pulsations can be likely resolved in this way. These questions are: can concurrent pulsation regimes exist? How does a changeover from one pulsation type to another occur? What does a chaotic behavior that is observed for many pulsating stars trigger? And so on. For these purposes, one-zone models are used. In these models, a star, which is a distributed system, is substituted for one homogenous (often, infinitely thin) layer that possess properties of a whole star in some sense. This eliminates a stellar stratification but reduces a system of partial differential equations to that of ordinary ones.

Historically, the first model taking into consideration both opacity effects and thermonuclear ones was the model by Zhevakin [6]. But it is little-known and poorly explored up till now that, obviously, associated with that the original work of Zhevakin was not translated into English. In this model, a star was substituted for a sphere of an elastic weightless gas surrounding by an infinitely thin envelope in which all the stellar mass concentrated. The gas with envelope could expand and compress; in so doing, an opacity of the gas changed. In the gas, thermonuclear reactions of the proton-proton chain or the CNO-cycle proceeded.

In this work the author have taken a second look at the Zhevakin's model.

---

[*] A. Ishlinsky Institute for Problems in Mechanics, Moscow, Russia; *e-mail*: m-i-ivanov@mail.ru

# 1. FORMULATION OF THE PROBLEM

Let consider a sphere of an elastic and weightless gas that is surrounded by an envelope of a mass $M$, which is in own gravity's field. Then, the equation of motion is:

$$M \frac{d^2 R}{dt^2} = -G \frac{M^2}{R^2} + 4\pi R^2 p \qquad (1.1)$$

where $R$ is a radial coordinate of the envelope, $t$ is time, $G$ is the gravitational constant, and $p$ is a pressure that is applied by the elastic gas to the envelope. Since we are constructing a one-zone model we will thereafter write ordinary derivatives instead of partial ones. The gas is supposed to be ideal (we neglect a radiation pressure):

$$p = R_g \rho T \qquad (1.2)$$

where $R_g$ is the gas constant, $T$ is a temperature of the gas, and $\rho$ is a density of the gas for which we have:

$$\rho = \frac{3M}{4\pi R^3} \qquad (1.3)$$

Thus, we can rewrite the equation (1.1) in the form:

$$\frac{d^2 R}{dt^2} = -G \frac{M}{R^2} + \frac{3 R_g T}{R} \qquad (1.4)$$

The second equation can be deduced from the first law of thermodynamics. The gas in the interior of the envelope is considered to show the infinitely high specific heat; that is, it is spatially-isothermal. Then, we obtain:

$$\frac{L}{M} - \varepsilon + p \frac{d}{dt}\left(\frac{1}{\rho}\right) + \frac{dU}{dt} = 0 \qquad (1.5)$$

where $L$ is a luminosity function, $\varepsilon$ is a thermonuclear energy generation rate, and $U$ is a specific internal energy of the gas for which we have:

$$U = \frac{1}{\gamma - 1} \frac{p}{\rho} \qquad (1.6)$$

where $\gamma = C_p / C_V$ is the ratio of the specific heats.

Substituting (1.2), (1.3), and (1.6) into (1.5), we find:

$$\frac{L}{M} - \varepsilon + 3 R_g \frac{T}{R} \frac{dR}{dt} + \frac{1}{\gamma - 1} R_g \frac{dT}{dt} = 0 \qquad (1.7)$$

The equations (1.4) and (1.7) form the system desired but the functions $L$ and $\varepsilon$ remain to be expressed.

# 2. LUMINOSITY FUNCTION

Based on isotropy of the radiation field inside the star ($\kappa \rho R_0 \gg 1$, where $R_0$ is a typical radius) we have the system for an energy density $E$, a radiation flux $H$ and a radiation pressure $P_R$ [7]:

$$\frac{dH}{dR} + \frac{2}{R} H + c \kappa \rho E - j \rho = 0 \qquad (2.1)$$

$$\frac{dP_R}{dR} + \frac{\kappa \rho}{c} H = 0 \qquad (2.2)$$

$$P_R = \frac{E}{3} \qquad (2.3)$$

where $c$ is the speed of light, $\kappa$ is an opacity, $j$ is an emissive power which is a sum of a heat radiation and the thermonuclear energy generation rate $\varepsilon$:

$$j = \kappa a c T^4 + \varepsilon \qquad (2.4)$$

where $a$ is the Stefan-Boltzmann constant.

For the luminosity function, we have the geometric relation [7]:

$$L = 4\pi R^2 H \qquad (2.5)$$

Let consider the system (2.1)-(2.3) in an infinitely thin stellar layer. With (2.4) and (2.5), we obtain:

$$\frac{L}{M} + c\kappa \left( E - aT^4 \right) - \varepsilon = 0 \qquad (2.6)$$

$$E - \frac{\kappa}{c} \frac{9M}{\left(4\pi R^2\right)^2} L = 0 \qquad (2.7)$$

Then, we express $\varepsilon$ from the equation (1.7) and substitute it into the formula (2.6):

$$E - aT^4 - \frac{R_g}{c\kappa} \left( 3\frac{T}{r}\frac{dR}{dt} + \frac{1}{\gamma - 1}\frac{dT}{dt} \right) = 0 \qquad (2.8)$$

From (2.7) and (2.8), we find:

$$L = \frac{\left(4\pi R^2\right)^2}{9M} \frac{1}{\kappa} \left\{ acT^4 + \frac{R_g}{\kappa} \left( 3\frac{T}{R}\frac{dR}{dt} + \frac{1}{\gamma - 1}\frac{dT}{dt} \right) \right\} \qquad (2.9)$$

Hence, if the opacity function $\kappa$ will be known, the formula (2.9) gives the required expression for the luminosity function. We will thereafter use the opacity function in the form:

$$\kappa = \kappa_0 \rho T^{-7/2} \qquad (2.10)$$

where $\kappa_0$ is some dimensional constant, in a similar fashion to [7][1]. Hence, with (2.10), one may rewrite the formula (2.9) in the form:

$$L = \frac{(4\pi)^3}{27\kappa_0 M^2} R^7 T^{7/2} \left\{ acT^4 + \frac{4\pi R_g}{\kappa_0 M} \left( R^2 T \frac{dR}{dt} + \frac{1}{3(\gamma - 1)} R^3 \frac{dT}{dt} \right) \right\} \qquad (2.11)$$

Let enter a dimensionless radial coordinate $r = R/R_0$, a temperature $\tau = T/T_0$, where is $T_0$ is a typical temperature, and a time $\tilde{t} = t/\Omega$, where $\Omega$ is a typical time, in (2.11):

$$L = \frac{(4\pi)^3}{27 M^2} \frac{ac}{\kappa_0} R_0^7 T_0^{15/2} r^6 \tau^{7/2} \left\{ r\tau^4 + \frac{4\pi}{ac} \frac{R_g R_0^3 T_0^{1/2}}{\kappa_0 M \Omega} r^3 \tau^{7/2} \left( \tau\dot{r} + \frac{1}{3(\gamma - 1)} r\dot{\tau} \right) \right\} \qquad (2.12)$$

where differentiation with respect to $\tilde{t}$ is denoted by dot.

For us, to derive the dimensional constant $\kappa_0$ is of major difficulty. Examination of classical Cepheid model opacities from [8] shows that the formula (2.10) is approximately obeyed for them and we may roughly estimate $\kappa_0 \sim 10^{21}$ kg$^{-2}$m$^5$K$^{-3.5}$ (see Fig. 1). Hence, for reasonable values of $R_0$, $M$, $T_0$, and $\Omega$, it is easy to see[2]:

$$\frac{4\pi}{ac} \frac{R_g R_0^3 T_0^{1/2}}{\kappa_0 M \Omega} \ll 1 \qquad (2.13)$$

Thus, with (2.13), the formula (2.12) can be rewritten in the form[3]:

---

[1] Zhevakin used more general law $\kappa = \kappa_0 \rho^n T^{-s}$, where $n$ and $s$ is some degrees, in the original work [6].

[2] One can see that the condition $\kappa \rho R_0 \gg 1$ likewise is true under circumstances considered.

[3] This deduction was absent in the Zhevakin's original work [6]. He used a formula of type of (2.14) immediately.

$$L = \frac{(4\pi)^3}{27} \frac{ac}{\kappa_0 M^2} R_0^7 T_0^{15/2} r^7 \tau^{15/2} \tag{2.14}$$

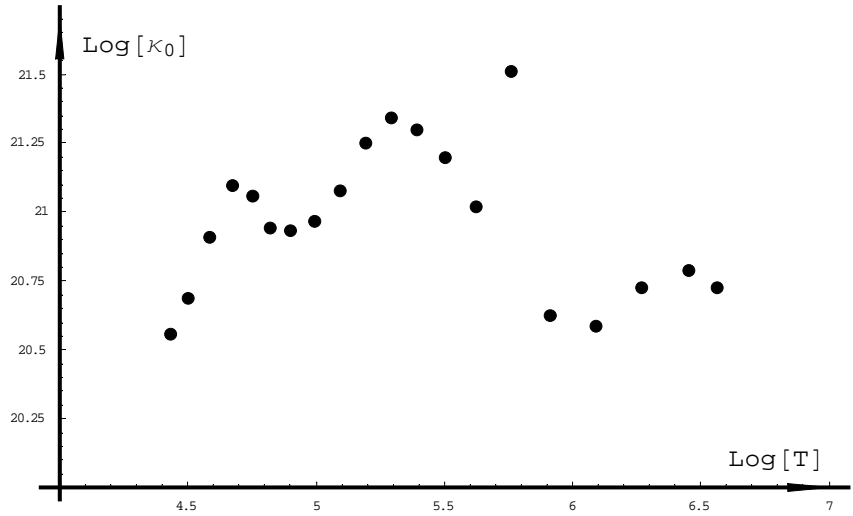

**Fig. 1.**

## 3. THERMONUCLEAR ENERGY GENERATION RATE

In real stars, different thermonuclear fusion processes occur. For a main-sequence star, the basic processes are the proton-proton chain and the CNO-cycle; the first one prevails at lower temperatures, whereas the second one prevails at higher temperatures. But in Cepheids that leave the main sequence and are located in the blue loop, the thermonuclear reactions of the triple-alpha process occur [2, 9]. Note that the triple-alpha process was not considered in the original Zhevakin's work but the proton-proton chain or the CNO cycle was [6]. The reactions of the triple-alpha process are [7]:

$He^4 + He^4 \rightleftarrows Be^8 + \gamma - 0.095\,MeV$

$Be^8 + He^4 \rightarrow C^{12} + \gamma + 7.4\,MeV$

For its energy generation rate, the formula is known [10][4]:

$$\varepsilon_{3\alpha} = 3.9 \cdot 10^{15} f \rho^2 Y^3 T^{-3} \exp\left\{-\frac{4.294 \cdot 10^9}{T}\right\} \tag{3.1}$$

where $f$ is a (dimensionless) screening factor depending on the occurrence of the $C^{12} + He^4$-reaction [15], $Y$ is a mass fraction of the helium in the star. The dimension of $\varepsilon_{3\alpha}$ is Wt/kg.

In the dimensionless variables, the formula (3.1) is rewritten in the form:

$$\varepsilon_{3\alpha} = 2.2 \cdot 10^{14} \frac{fM^2 Y^3}{R_0^6 T_0^3} r^{-6} \tau^{-3} \exp\left\{-\frac{4.294 \cdot 10^9}{T_0} \tau^{-1}\right\} \tag{3.2}$$

---

[4] We use the traditional formula for the triple-alpha reaction rate. Ogata et al. [11] has obtained new values for the triple-alpha reaction rate that are vastly more than the traditional ones for temperatures in the area of $10^7 - 10^8$ K. This leads to revolutionary astrophysical consequences, among which are the shortening or disappearance of the red giant phase for low-mass stars [12], the absense of helium flashes for low- and intermediate-mass stars [13] and the uniqueness of the instability strip crossing by Cepheids [14]. The latter implies that a Cepheid's period must only increases with time but never not decreases [14]. All these consequences contradict observations, then, the author supposes that the result of [11] calls for confirmation.

## 4. EXPLICIT FORM SYSTEM AND DIMENSIONLESS NUMBERS

The typical time $\Omega$ can be expressed by the typical radial coordinate $R_0$, the stellar mass $M$ and the gravitational constant $G$. Let use the form of [16, 17]:

$$\Omega = \sqrt{\frac{R_0^3}{\gamma GM}} \tag{4.1}$$

With (4.1), let substitute the formulae (2.14) and (3.2) into the system (1.4), (1.7) and turn to dimensionless variables. We obtain:

$$\gamma r^2 \ddot{r} + 1 = 3 \frac{R_g T_0 R_0}{GM} r\tau \tag{4.2}$$

$$1.66 \cdot 10^{-5} \frac{R_0^{17/2} T_0^{13/2}}{\kappa_0 R_g G^{1/2} M^{7/2}} r^8 \tau^{15/2} -$$

$$-2.2 \cdot 10^{14} \frac{fM^{3/2} Y^3}{R_g G^{1/2} R_0^{9/2} T_0^4} r^{-5} \tau^{-3} \exp\left\{-\frac{4.294 \cdot 10^9}{T_0} \tau^{-1}\right\} + 3\sqrt{\gamma}\tau\dot{r} + \frac{\sqrt{\gamma}}{\gamma - 1} r\dot{\tau} = 0 \tag{4.3}$$

where the known numerical values of $a$ and $c$ are used.

Thus, we find the dimensionless numbers:

$$A = \frac{R_g T_0 R_0}{GM} \tag{4.4}$$

$$B = 1.66 \cdot 10^{-5} \frac{R_0^{17/2} T_0^{13/2}}{\kappa_0 R_g G^{1/2} M^{7/2}} \tag{4.5}$$

$$C = 2.2 \cdot 10^{14} \frac{fM^{3/2} Y^3}{R_g G^{1/2} R_0^{9/2} T_0^4} \tag{4.6}$$

$$\lambda = \frac{4.294 \cdot 10^9}{T_0} \tag{4.7}$$

Then, the system (4.2)-(4.3) can be rewritten in the form:

$$\gamma r^2 \ddot{r} - 3Ar\tau + 1 = 0 \tag{4.8}$$

$$3\sqrt{\gamma}\tau\dot{r} + \frac{\sqrt{\gamma}}{\gamma - 1} r\dot{\tau} + Br^8 \tau^{15/2} - Cr^{-5} \tau^{-3} \exp\{-\lambda\tau^{-1}\} = 0 \tag{4.9}$$

The system (4.8)-(4.9) is the system of ODE's of the total order 3. To gain a better insight into this system, we need find out what values do the dimensionless numbers $A$, $B$, $C$, and $\lambda$ assume? It is worth noting that our interest here is only with crude estimations of the dimensionless numbers (4.4)-(4.7).

For $R_g$, we have the formula $R_g = R_u / \mu$, where $R_u = 8.31 \text{ J}/(\text{mol} \cdot \text{K})$ is the universal gas constant, and $\mu$ is the average molecular mass of the substance of the star for which we have the formula of [7]:

$$\frac{1}{\mu} = \left(2X + \frac{3}{4}Y + \frac{1}{2}Z\right) \cdot 10^3 \tag{4.10}$$

where $X$ is a mass fraction of the hydrogen in the star, $Z$ is a mass fraction of the heavy elements, and the factor $10^3$ appears in going from grams to kilograms.

From (4.10), we have $\mu = (4/3) \cdot 10^{-3}$ kg/mol for an all-helium layer, whereas for a layer of the standard abundance of a main-sequence Population I star $X = 0.7$, $Y = 0.28$, $Z = 0.02$ [18], it is easy to obtain $\mu = 0.62 \cdot 10^{-3}$ kg/mol. Since the gas constant enters into both formulae (4.4) and (4.5) to the 1 power one may use $R_g \sim 10^4 \text{ J}/(\text{kg} \cdot \text{K})$ as a crude approximation.

For $\kappa_0$, we have found the estimation $\kappa_0 \sim 10^{21}$ kg$^{-2}$m$^5$K$^{-3.5}$ in the Section 2. Substantial uncertainties occur for the screening factor $f$ since it is hard to tell whether how is the rate of the $C^{12} + He^4$-reaction in Cepheids? This reaction is supposed to be of importance for rather massive stars ($M > 10 M_\odot$, crudely) [19]. But the screening factor enters into the formula (4.6) to the 1 power, then, one may use $f \sim 1$ as a crude approximation. The layer of thermonuclear energy generation is considered to be a near-all-helium, thus, we suppose that $Y^3 \sim 1$.

The main weakness of the Zhevakin's model is that radiation confinement and thermonuclear energy generation are properly site of one layer in the Zhevakin's model, which is not the case, obviously. As a consequence, in order to estimate magnitudes of dimensionless numbers it will to bear in mind that we cannot use common values of $M$, $R_0$, and $T_0$ to both sites.

To estimate $B$, we need consider physical conditions of the layer where stellar radiation confines within. It is known that the radiation confinement site is the He II ionization zone with typical temperatures in the order of $(3.5 - 5.5) \cdot 10^4$ K [1]. Due to low temperature this zone must lie in the close proximity of the stellar surface [2], that is, typical radius is a stellar one that can be found by the Baade-Wesselink method. It was determined that Cepheids have large radii of the order of $50 - 100 R_\odot$ [20-23]. Thus, we can take $R_0 \sim 5 \cdot 10^9$ km, crudely.

From the stellar evolution theory, Cepheids is supposed to have masses in the interval $3 - 12 M_\odot$ [24]. We consider the layer where stellar radiation confines within by the opacity (kappa) mechanism. Since outer layers of the star are low-massive, therefore, it is believed that $M \sim 5 M_\odot \approx 10^{31}$ kg.

Substituting these values into the formula (4.5), we obtain $B \sim 10^{-12} - 10^{-13}$.

Let estimate $C$. The triple-alpha reaction occurs at a temperature of the order of $10^8$ K [7]. Then, it is conceivable that $T_0 \sim 10^8$ K.

Let rearrange the formula (4.6) to the form:

$$C = 1.9 \cdot 10^{15} \frac{f \langle \rho \rangle^{3/2} Y^3}{R_g G^{1/2} T_0^4} \qquad (4.11)$$

where $\langle \rho \rangle$ is the average density of the helium-burning region of the star.

Estimation of $\langle \rho \rangle$ involves great difficulties. Because of this, for simplicity we use the central density of the star $\rho_C$ having regard to the obvious inequality $\langle \rho \rangle < \rho_C$. For the central density of a $5 M_\odot$ Cepheid at the second crossing of the instability strip, we have the estimation $\rho_C \sim 10^7$ kg/m$^3$ [25]. For more massive Cepheids, the central density can be as great as something like $10^8$ kg/m$^3$ [26]. Thus, we take $\langle \rho \rangle < 10^8$ kg/m$^3$.

Substituting these values into the formula (4.11), we obtain $C < 2 \cdot 10^{-4}$.

Smallness of $B$ and $C$ is just a consequence of the well-known fact that the thermonuclear heating and the radiation cooling forces fall far short of the gravity or the adiabatic compression/expression ones [6].

## 5. ZEROTH APPROXIMATION

Let denote $C = \delta$ and $B = \Lambda \delta$, where $\Lambda \sim 1$. Then, the equation (4.9) is rewritten in the form:

$$3\sqrt{\gamma}\tau\dot{r} + \frac{\sqrt{\gamma}}{\gamma-1}r\dot{\tau} + \delta\left(\Lambda r^8\tau^{15/2} - r^{-5}\tau^{-3}\exp\{-\lambda\tau^{-1}\}\right) = 0 \qquad (5.1)$$

Let search for a solution of the system (4.8), (5.1) in the form:
$$r(t) = r_0(t) + \delta r_1(t) + \delta^2 r_2(t) + ... \qquad (5.2)$$
$$\tau(t) = \tau_0(t) + \delta\tau_1(t) + \delta^2\tau_2(t) + ... \qquad (5.3)$$

Substitute the expansions (5.2) and (5.3) into the system (4.8), (5.1). Then, we obtain:
$$\gamma r_0^2 \ddot{r}_0 - 3A r_0 \tau_0 + 1 + \delta\left(\gamma r_0^2 \ddot{r}_1 + 2\gamma r_0 r_1 \ddot{r}_0 - 3A r_1 \tau_0 - 3A r_0 \tau_1\right) + O(\delta^2) = 0 \qquad (5.4)$$

$$3\sqrt{\gamma}\tau_0 \dot{r}_0 + \frac{\sqrt{\gamma}}{\gamma-1}r_0 \dot{\tau}_0 +$$
$$+\delta\left(3\sqrt{\gamma}\tau_1 \dot{r}_0 + 3\sqrt{\gamma}\tau_0 \dot{r}_1 + \frac{\sqrt{\gamma}}{\gamma-1}r_1 \dot{\tau}_0 + \frac{\sqrt{\gamma}}{\gamma-1}r_0 \dot{\tau}_1 + \Lambda r_0^8 \tau_0^{15/2} - r_0^{-5}\tau_0^{-3}\exp\{\lambda\tau_0^{-1}\}\right) + \qquad (5.5)$$
$$+O(\delta^2) = 0$$

Thus, keeping only terms of the order of $O(1)$ in the system (5.4)-(5.5) we derive the zeroth approximation system:
$$\gamma r_0^2 \ddot{r}_0 - 3A r_0 \tau_0 + 1 = 0 \qquad (5.6)$$
$$3(\gamma - 1)\tau_0 \dot{r}_0 + r_0 \dot{\tau}_0 = 0 \qquad (5.7)$$

The equation (5.7) is solvable algebraically; the exact solution can be rewritten in the form:
$$\tau_0 = C_1 r_0^{3(1-\gamma)} \qquad (5.8)$$
where $C_1$ is an integration constant. Substituting (5.8) into the equation (5.6), we obtain the well-known equation of Bhatnagar and Kothari [17, 27, 28]:
$$\gamma r_0^2 \ddot{r}_0 - C_2 r_0^{4-3\gamma} + 1 = 0 \qquad (5.9)$$
where $C_2 = 3AC_1$ is some constant. The equation (5.9) defines adiabatic hydrodynamic approximation of the Cepheid pulsation problem; for we suppose both the luminosity function and the thermonuclear energy generation rate to be small this result is obvious.

The equation (5.9) has to be supplemented by two initial conditions:
$$r_0(0) = 1 \qquad (5.10)$$
$$r_0'(0) = 0 \qquad (5.11)$$

For some $C_2$, the Cauchy problem (5.9)-(5.11) has periodic solutions [17, 27-29].
For the luminosity function, we obtain the formula, substituting (5.8) into (2.14)[5]:
$$L = L_0 r_0^{(59-45\gamma)/2} \qquad (5.12)$$
where $L_0$ is some constant which value is not essential for us now.

Close inspection of the problem (5.9)-(5.11) shows that radius-time curves $r_0(t)$ are symmetrical about mid-period points. Typical curves are plotted in the Fig. 2.

Consequently, luminosity curves $L(t)$ defining by the formula (5.12) are symmetrical about mid-period points, too. The luminosity curves associated with the Fig. 2 are plotted in the Fig. 3. But real Cepheids show asymmetric luminosity curves [1]. In addition, the luminosity maximum is often not synchronized with the radius minimum for real Cepheids [2].

---

[5] It is worth noting that no relation for the luminosity function can find in the framework of pure hydrodynamic models.

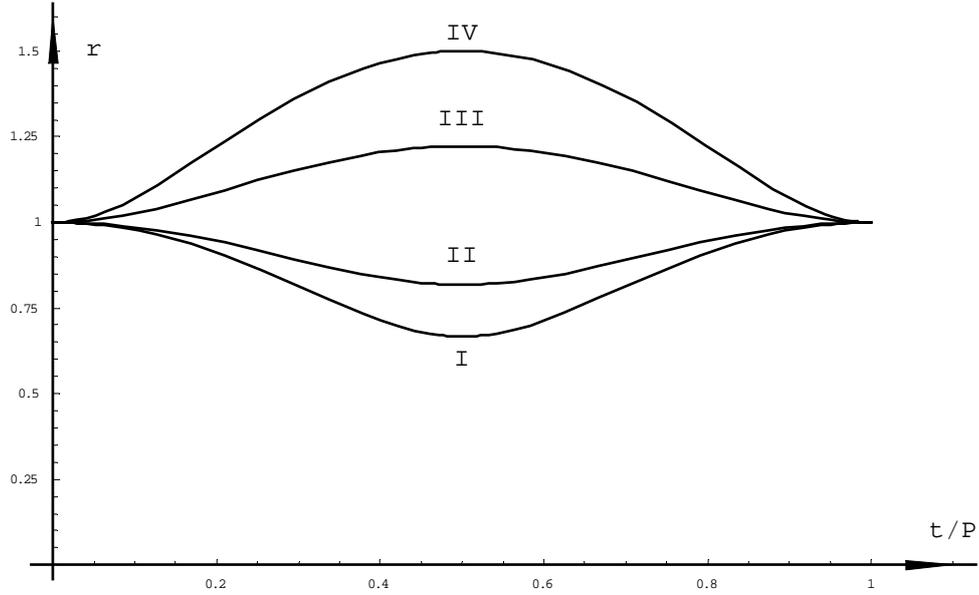

**Fig. 2.** $\gamma = 5/3$ for all the curves. Constants $C_2$ and (dimensionless) pulsation periods $P$: I − $C_2 = 0.8$, $P = 6.171$, II − $C_2 = 0.9$, $P = 7.031$, III − $C_2 = 1.1$, $P = 9.500$, IV − $C_2 = 1.2$, $P = 11.336$.

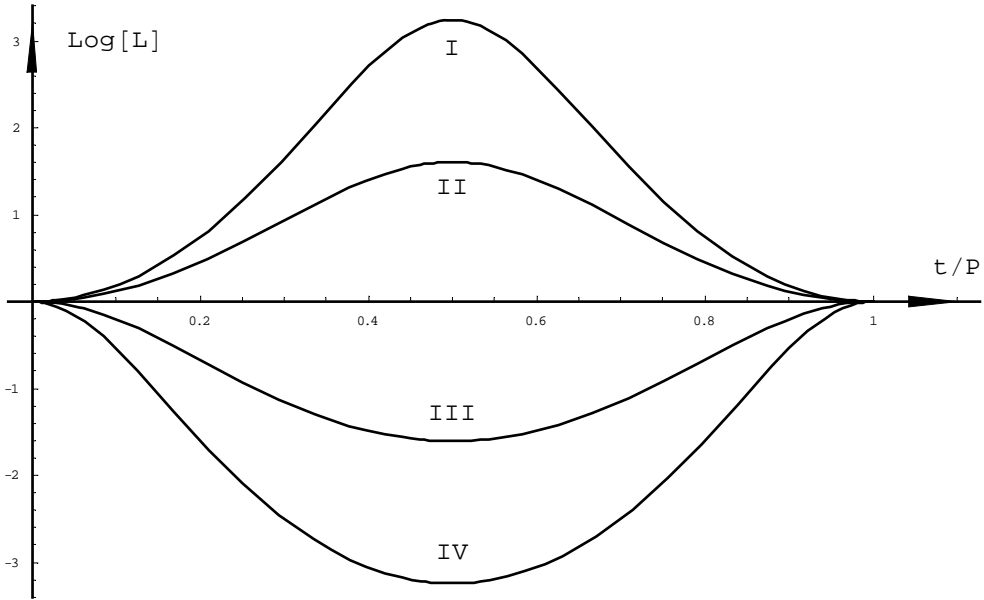

**Fig. 3.** $\gamma = 5/3$ and $L_0 = 1$ for all the curves. Constants $C_2$ and (dimensionless) pulsation periods $P$: I − $C_2 = 0.8$, $P = 6.171$, II − $C_2 = 0.9$, $P = 7.031$, III − $C_2 = 1.1$, $P = 9.500$, IV − $C_2 = 1.2$, $P = 11.336$.

## CONCLUSION

From the Section 4 we know that $\delta$ is very small. This leads to the fact that one has to take account of the "luminosity" and "thermonuclear" terms of the equation (1.5) only in some superficial layer where the adiabatic terms are likewise small. In the stellar interior the dynamics of the star is dictated by balance of the gravity force and the pressure. This result correlates with the fact that the adiabatic theory gives a good approximation for non-adiabatic pulsation periods [24].

The Zhevakin's model has no superficial layer since the star is supposed to be spatially-isothermal; it, in fact, reduces to the equation of Bhatnagar and Kothari. Because of smallness of actual $\delta$, consideration of terms of the first order of smallness does not involve principal changes of the results of the Section 5. Thus, such phenomena as asymmetry of luminosity curves, phase shifts, the Hertzsprung progression cannot be understood in the framework of the model considered.

Does a particular one-zone model describe real Cepheid dynamics? J.P. Cox speculated that, perhaps, it was not [30]. On the other hand, Zhevakin supposed that one-zone models could approximate fundamental tones of pulsations of stars with stratified structures [6]. But, since one has to take into consideration non-hydrodynamic terms only near the stellar surface two-zone models seem to be of more interest. Besides, it is unlikely that such simplified models as one- (or even two-) zone ones allow to find adequate pulsation periods. From the above discussion it is clear that the adiabatic periods give the approximation needed but the stellar stratification must be kept in mind. It can be done, for instance, in such a manner as in [17] where it was assumed that deviations from the equlibrium were small and there were no energy flux between stellar layers as a star pulsated; later condition corresponded to the fundamental tone pulsation.